\documentclass[conference,a4paper]{IEEEtran}
\IEEEoverridecommandlockouts
\input{common_def.sty}

\author{
  \IEEEauthorblockA{Mingchao Yu \quad Parastoo Sadeghi \quad Neda Aboutorab}\\\vspace{-5pt}
  \IEEEauthorblockN{\small Research School of Engineering, The Australian National University, Canberra, Australia \\ \texttt{\{ming.yu, parastoo.sadeghi, neda.aboutorab\}@anu.edu.au} \\\vspace{-2em}
  }}
\title{On Deterministic Linear Network Coded Broadcast and Its Relation to Matroid Theory}
\begin{document}
\sloppy

\maketitle
\begin{abstract}
Deterministic linear network coding (DLNC) is an important family of network coding techniques for wireless packet broadcast. In this paper, we show that DLNC is strongly related to and can be effectively studied using matroid theory without bridging index coding. We prove the equivalence between the DLNC solution and matrix matroid. We use this equivalence to study the performance limits of DLNC in terms of the number of transmissions and its dependence on the finite field size. Specifically, we derive the sufficient and necessary condition for the existence of \emph{perfect} DLNC solutions and prove that such solutions may not exist over certain finite fields. We then show that identifying perfect solutions over any finite field is still an open problem in general. To fill this gap, we develop a heuristic algorithm which employs graphic matroids to find perfect DLNC solutions over any finite field. Numerical results show that its performance in terms of minimum number of transmissions is close to the lower bound, and is better than random linear network coding when the field size is not so large.

\end{abstract}

\section{Introduction}

The broadcast nature of wireless medium allows a sender to simultaneously serve multiple receivers who are interested in the same information. A basic wireless broadcast problem is how to efficiently broadcast a block of data packets to a set of wireless users subject to packet erasures using the minimum number of transmissions \cite{keller:drinea:fragouli:2008,nguyen:tran:nguyen:bose:2009,medard:2008:3receivers,sprintson:min:2007}.

A desirable solution to this problem is network coding \cite{keller:drinea:fragouli:2008,nguyen:tran:nguyen:bose:2009,medard:2008:3receivers}, which allows the sender to code data packets together and send the coded packets. It can substantially save the number of transmissions compared with uncoded packet scheduling \cite{keller:drinea:fragouli:2008,nguyen:tran:nguyen:bose:2009}.
It can even achieve throughput optimality, namely, every successfully received coded packet brings new information to the receivers that are missing packets.
When coded packets are linear combinations of the data packets with coefficients chosen from a finite field, the technique is known as linear network coding \cite{Yeung:linearNC:2003}.

The choice of coding coefficients divides linear network coding techniques into two classes: random linear network coding (RLNC) and deterministic linear network coding (DLNC). RLNC, in which coding coefficients are randomly chosen, is asymptotically throughput optimal at the price of high decoding computational complexity \cite{heide_systematic_RLNC} and large packet decoding delay \cite{yu:neda:parastoo:isit2013}. On the other hand, DLNC techniques offer merits such as low decoding computational complexity and small packet decoding delay, but require more complicated coding process at the sender.

In DLNC techniques, the sender first sends the data packets in a block uncoded once. The receivers will receive subsets of the data packets due to packet erasures. Based on this reception instance, the sender generates coded packets, which are referred to as a solution, by determining which data packets to code together, which field size to use, and what the coefficients should be. The reception instance is updated at an appropriate frequency by collecting feedback from receivers.

There are various DLNC techniques in the literature with some interesting results, such as opportunistic broadcast \cite{keller:drinea:fragouli:2008}, instantly decodable network coding \cite{sadeghi:adaptive_broadcast_2009,sameh:valaee:globecom:2010,yu:parastoo:neda:idnc2013}, and sparse innovative coding \cite{2011_Kwan_sparse}.
However, some key knowledge gaps still exist for DLNC. Specifically, the following three questions have not been fully addressed:

\begin{enumerate}[\textbf{\emph{Q-}}1]
\item What is the minimum number of coded transmissions using DLNC to complete the broadcast of a block of data packets?
\item How does this number depend on the field size?
\item How can this number be achieved?
\end{enumerate}
While the answer to \emph{Q}-1 reveals the performance limits of DLNC, \emph{Q}-2 is motivated by the desire to reduce the field size and hence the computational complexity, and \emph{Q}-3 is related to the design of coding algorithms. In this paper, we will answer these questions by using matroid theory.

\subsection{Related works}
\subsubsection{Matroid theory} Matroid theory is a branch of mathematics capturing and generalizing linear independence in vector space \cite{Matroid_theory}. A matroid $M$ is represented by a pair $(E,\I)$, where $E$ is a finite set of elements and $\I$ is a family of subsets of $E$ called independent sets. Its common representations include matrix matroid and graphic matroid.

The relation between matroid theory and the general network coding problem has been well studied \cite{dougherty:nwmatroid:2007,sun:nwmatroid:2008,dougherty:nwmatroid:2010}. But its extension to the DLNC problem for wireless broadcast has not been addressed. Indeed, DLNC is only indirectly and partially associated to matroid theory through index coding.

\subsubsection{Index coding}
Index coding \cite{sprintson:algorithm:2008,sprintson:ic_nc_matroid:2010,Yossef:index:2011,effros:equivalence:isit2013} can be viewed as a special DLNC technique \cite{sprintson:min:2007}. It also designs solutions for instances of partially received data blocks. It is special because in index coding 1) every (virtual) receiver wants only one data packet. Hence, once a DLNC problem is converted to an index coding problem, the information about what packets \emph{real} receivers are missing is lost; and 2) transmissions are erasure free. Its relation to the general network coding problem has been discussed in \cite{sprintson:ic_nc_matroid:2010,effros:equivalence:isit2013}.

Properties of index coding have been studied using matroid theory. The authors of \cite{sprintson:min:2007} showed through a matroid example that the minimum number of coded transmissions of index coding does not necessarily decrease monotonically with increasing the field size. Later in \cite{sprintson:ic_nc_matroid:2010}, they proved the equivalence between the perfect solution of index coding (which will be defined in the next subsection) and a matroid.

\begin{figure}[t]
\centering
\subfigure[The relation between DLNC for wireless broadcast and matroid theory, bridged by index coding.]{\includegraphics[width=0.45\linewidth]{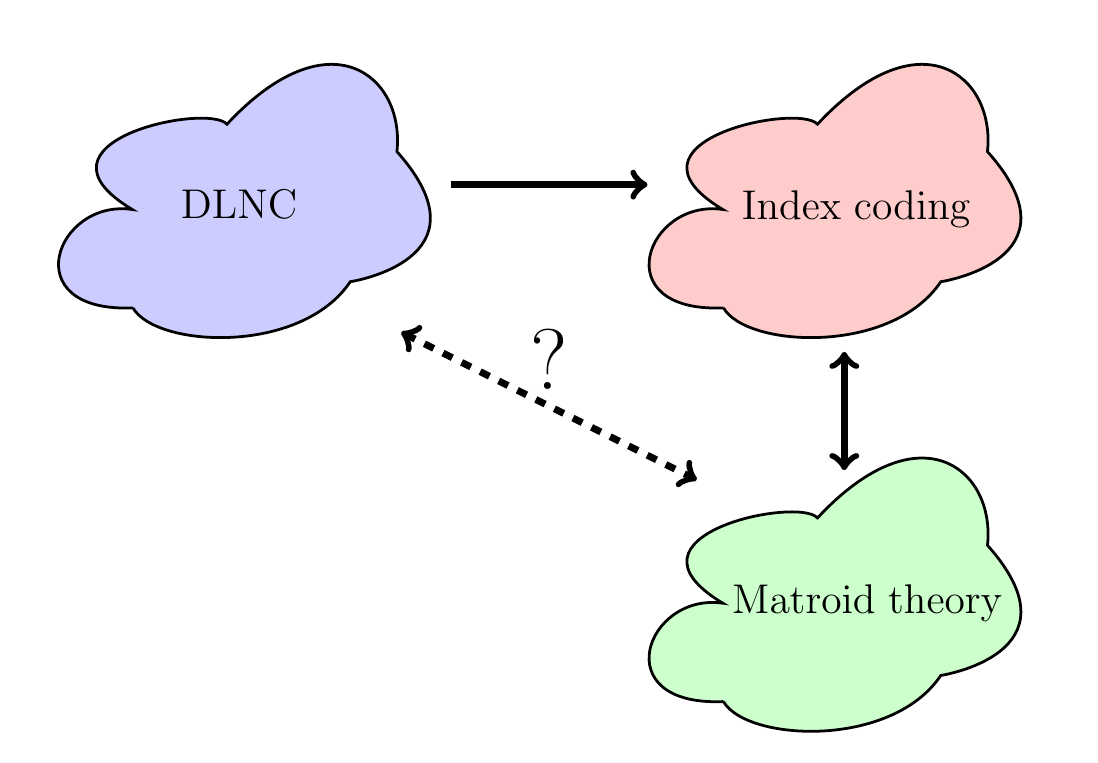}}\hspace{5pt}
\subfigure[The direct relation we construct between DLNC for wireless broadcast and matroid theory.]{\includegraphics[width=0.51\linewidth]{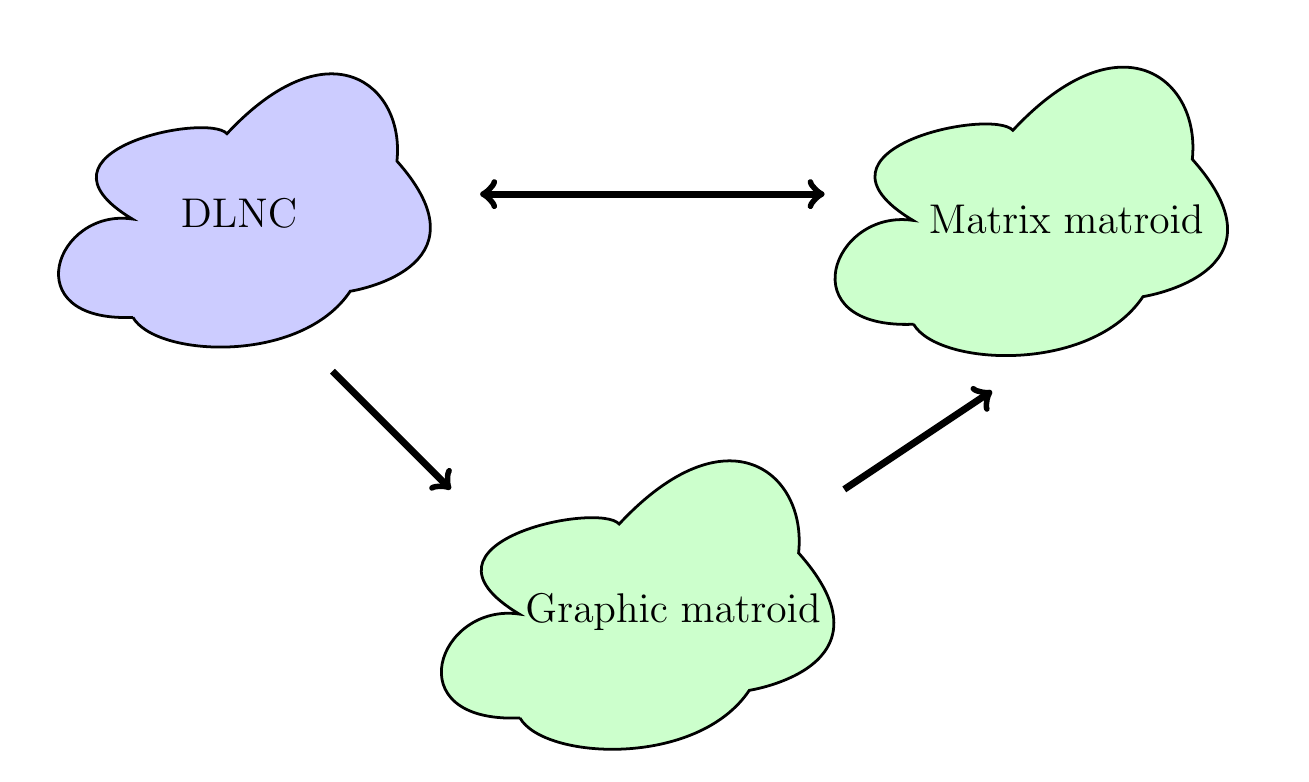}}
\caption{The relation between DLNC for wireless broadcast and other areas.}
\label{fig:topics}
\end{figure}

However, such equivalence has some limitations that prevent the full usage and interpretation of results in matroid theory to index coding, and consequently to DLNC. First, the matroid can only be represented if the perfect index coding solution is known in advance. Second, the number of data packets is larger than the number of elements in the matroid, and thus there is no bijection between the two. Finally, the packet reception instance is specially designed and thus cannot be generalized.  A sketch of the relation among these three areas is shown in \figref{fig:topics}(a).

\subsection{Contributions}

In this work we construct a direct relation between DLNC and matroid theory without relying on index coding. This construction motivates many fundamental research opportunities of DLNC by using matroid theory. Among them, we answer the aforementioned three questions and provide the following contributions:
\begin{enumerate}
\item We prove the equivalence between the DLNC solution and matrix matroid;
\item We propose sufficient and necessary conditions for the existence of the perfect DLNC solutions, which can achieve the lower bound on the minimum number of coded transmissions. We then prove that a perfect solution may not exist over certain finite fields. We also show that a complete answer to \emph{Q}-2 is still unknown. Some of our results revisit and refine those in \cite{sprintson:min:2007}.
\item Motivated by the above limitations, we develop a DLNC algorithm which can produce DLNC solutions for any packet reception instance over any finite field by heuristically finding a graphic matroid and then a matrix matroid. We show that its minimum number of coded transmissions performance is very close to the lower bound even when the field size is as small as 2.
\end{enumerate}

\section{System Model}
\subsection{Transmission Setup}
We consider packet-based wireless broadcast from one
sender to $N$ receivers. Receiver $n$ is denoted by $R_n$ and the
set of all receivers is  $\R = \{R_1, \cdots, R_{N}\}$. There are a
total of $K$ data packets with identical lengths to be delivered to all receivers. Packet $k$
is denoted by $\p_k$ and the set of all packets is  $\P = \{\p_1,
\cdots, \p_K\}$. Time is slotted
and in each time slot, one (coded or original data)  packet is
broadcast. The wireless channel between the sender and each receiver
is subject to independent memoryless packet erasures. 

\subsection{Systematic Transmission Phase and Receiver Feedback}

Initially, the $K$ data packets are transmitted uncoded once using $K$ time slots, constituting a \emph{systematic transmission phase} \cite{sameh:valaee:globecom:2010,yu:parastoo:neda:idnc2013}. After that, each receiver provides feedback to the sender about the packets it has received or lost.\footnote{We assume that there exists an error-free feedback link between each receiver to the sender that can be used with appropriate frequency.}  The complete states of receivers and packets, referred to as a (packet) reception instance, can be captured by an $N \times K$ state feedback matrix (SFM)  $\mA$ \cite{sameh:valaee:globecom:2010,yu:parastoo:neda:idnc2013}, where the element at row $n$ and column $k$ is denoted by $a_{n,k}$. We let $a_{n,k}=1$ if $R_n$ has lost $\p_k$ and let $a_{n,k}=0$ if $R_n$ has received $\p_k$.

Based on $\mA$, we define the \emph{Wants} set of each receiver \cite{sadeghi:adaptive_broadcast_2009}:

\begin{Definition}
The Wants set of receiver $R_n$, denoted by $W_n$, is a set of data packets which are lost at $R_n$ due to packet erasures. That is, $W_n = \{\p_k: a_{n,k} = 1\}$.
\end{Definition}

The size of $W_n$ is denoted by $w_n$. We further denote by $\wm$ the largest $w_n$ across all receivers. The collection of all Wants sets is denoted by $\W$. It suffices to describe a reception instance by $\W$.

\subsection{Coded Transmission Phase}

In this phase, a coded packet $\X_u$ is generated as:
\vspace{-0.4em}
\begin{equation}
\X_u=\sum_{k=1}^{K}c_{u,k}\p_k
\end{equation}
where $u$ is the time index, $\{c_{u,k}\}$ are linear coding coefficients chosen from a finite field $\F_q$. The coding coefficients of $U$ coded packets form a $U\times K$ coding matrix, which is denoted by $\mC$. Each row of $\mC$ is the coding vector $\c$ of a coded packet and is attached to that coded packet. We then have the notion of DLNC solution of any given reception instance:
\begin{Definition}
A coding matrix $\mC$ is a DLNC solution if:
\begin{enumerate}[{\textbf{S}-}1:]
\item The columns of $\mC$ indexed by $W_n$ are linearly independent for every receiver $R_n\in\R$. That is, $r(\mC(:,W_n))=w_n,~\forall n\in[1,N]$, where $r(\cdot)$ is the rank function;
\item Removing any single row from $\mC$ will violate \textbf{S}-1 for at least one receiver.
\end{enumerate}
\end{Definition}
\emph{S}-1 guarantees that, upon receiving all the coded packets that contain data packets in $W_n$, receiver $R_n$ can decode all its wanted data packets by solving linear equations. \emph{S}-2 implies that there is no redundant coding vector in $\mC$. Therefore, the number of rows, $U$, of a solution $\mC$ is its \emph{minimum number of coded transmissions} to satisfy the demands of all receivers. This minimum can be achieved in the presence of packet erasures with a non-zero probability.

Let $U_q$ be the smallest possible $U$ among all the solutions over $\F_q$. It is obvious that $U_q\nless\wm$, because $\wm$ is the information theoretical lower bound. Moreover, as we shall see later, there also exist reception instances in which $U_q>\wm$ for certain $q$. Therefore it can be concluded that $U_q\geqslant \wm$.

We then have the notions of optimal and perfect solutions:
\begin{Definition}
A solution $\mC$ over $\F_q$ is optimal if $U=U_q$.
A solution $\mC$ is perfect if $U=\wm$.
\end{Definition}

Finding the optimal/perfect solutions over a given $\F_q$ is of major interest in DLNC. It has a strong relation to matroid theory, which we now present.

\section{Preliminaries of Matroid Theory}

A matroid $M$ is an ordered pair $(E,\I)$. $E$ is a finite set of elements called the \emph{ground set}. $\I$ is a family of subsets of $E$ called \emph{independent sets}. An independent set is denoted by $I$ and its rank is equal to its cardinality, i.e., $r(I)=|I|$. A maximal independent set is called a basis \cite{Matroid_theory}. Every independent set of $M$ is a subset of one of the bases of $M$.



All subsets of $E$ not in $\I$ are called dependent sets. Among them, those whose ranks are one less than their cardinalities are called \emph{circuits}, denoted by $C$. That is, $r(C)=|C|-1$.

We now examine the meanings of independent set and circuit under two representations of matroid, including graphic matroid and matrix matroid, and then discuss their relation.

\subsection{Graphic matroid}
Let $\G(V,E)$ be an undirected graph with $|V|=U+1$ vertices, each denoted by $\v_i$ where $i\in[1,U+1]$, and $|E|=K$ edges, each denoted by $\e_j$ where $j\in[1,K]$. It defines a graphic matroid as follows:
\begin{Definition}
A graphic matroid $M(\G)$ has the edge set $E$ as its ground set. A subset $I$ of $E$ is an independent set if the edges in $I$ do not contain any graphic circuits.
\end{Definition}

The rank of a graphic matroid is $r(M(\G))=|V|-p$, where $p$ is the number of connected subgraphs\footnote{Any two vertices in a connected graph have at least one edge path connecting them.} of $\G$ \cite{Matroid_theory}. The rank of a connected graphic matroid is thus $|V|-1=U$.

\subsection{Matrix matroid}
Let $\mC$ be a $U\times K$ matrix of rank $U$ over $\F_q$. It defines a matrix matroid as follows:
\begin{Definition}
A matrix matroid $M(\mC)$ has the collection of all columns of $\mC$ as its ground set $E$. A subset $I$ of $E$ is an independent set if the columns in $I$ are linearly independent.
\end{Definition}
It is intuitive that the rank of the matroid $M(\mC)$ is $U$. Next, we define the notion of $q$-representability of a matroid:

\begin{Definition}
A matroid $M(E,\I)$ having $K=|E|$ elements and rank $r_M$ is called $q$-representable if there exists an $r_M\times K$ matrix $\mC$ over $\F_q$ without any all-zero columns such that its columns indexed by any $I\in\I$ are linearly independent.
\end{Definition}

\subsection{The relation between graphic and matrix matroids}

From now on, we will use $\e_i$ to denote both an edge in a graphic matroid and a column in a matrix matroid. The following lemma describes the relation between graphic matroid and matrix matroid:

\begin{Lemma}\label{lemma:graph_matrix}
A graphic matroid is $q$-representable for any $q$, but a matrix matroid may not have a graphic representation.
\end{Lemma}

We now demonstrate via an example the first half of this lemma, i.e.,  how to construct a matrix matroid from a graphic matroid. Examples of the second half of this lemma can be found in  \cite{Matroid_theory}.

\begin{figure*}[ht]
\centering
\includegraphics[width=0.75\linewidth]{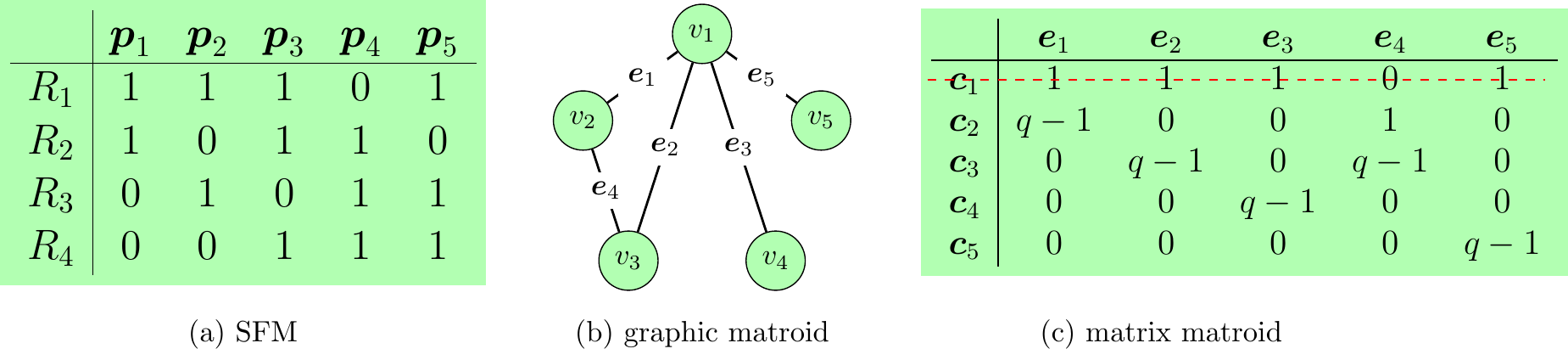}
\caption{Obtaining the matrix matroid/DLNC solution of a reception instance via graphic matroid.}
\label{fig:algorithm}
\end{figure*}

\begin{Example}\label{example:graph_matrix}
The graph $\G(V,E)$ in \figref{fig:algorithm}(b) defines a graphic matroid $M(\G)$ with five elements $E=\{\e_1,\cdots,\e_5\}$. The only circuit is $\{\e_1,\e_2,\e_4\}$. All the other proper subsets of $E$ are independent sets. The rank of $M(\G)$ is $|V|-1=4$.

We now generate a $|V|\times|E|$ all-zero matrix $\mC$. For every column $\e_i$, we assign $\{1,q-1\}$ to entries $\{c_{m,i},c_{n,i}\}$ if vertices $\v_m$ and $\v_n$ are incident to edge $\e_i$ and $m<n$,. Since every column contains exactly a pair of $\{1,q-1\}$, we can safely remove a single row from $\mC$ without losing any information. The $4\times5$ matrix shown in \figref{fig:algorithm}(c), after removing the first row, is the target matrix matroid. One can easily verify that the columns of $\mC$ have the same independency as the edges in the graphic matroid. Specifically, columns $\e_1,\e_2$, and $\e_4$ are dependent because $\e_1-\e_2+\e_4=\mathbf{0}$.

\end{Example}

\section{Connecting DLNC to Matroid Theory}

By comparing the notions of DLNC solution and matrix matroid, an equivalence between them can be constructed:
\begin{Theorem}
Given a reception instance with packet set $\P$ and the collection of Wants sets $\W$, its DLNC solution over $\F_q$ is equivalent to matrix matroid in a way that:
\begin{itemize}
\item Every $U\times K$ solution $\mC$ defines a matrix matroid $M(\mC)$ over $\F_q$ which has $K$ elements and a rank of $U$;
\item The matrix representation of any $q-$representable matroid with ground set $E=\P$ and $\I\supseteq\W$ is a solution.
\end{itemize}
\end{Theorem}

Hence, the $q$-representability of a matroid strongly couples with a basic question in DLNC: \emph{is there a way to identify the perfect DLNC solution over a given finite field?} We answer it positively by first deducing two corollaries of Theorem 1, and then showing that its general answer is still an open problem.

\begin{Corollary}
A sufficient condition for the existence of perfect solutions of a reception instance is that, the uniform matroid $U_{K}^{r}$ is $q$-representable, where $r=\wm$. This condition becomes both necessary and sufficient when $\W=\I$.
\end{Corollary}
Here a uniform matroid $U^r_K$ has $|E|=K$ elements. Every $r$-element subset of $E$ is a basis. The proof of this corollary relies on the fact that the $\I$ of $U_K^r$ is the superset of any $\W$ with $\wm=r$, and is omitted here due to the page limit.

\begin{Corollary}
The optimal solution of any reception instance exists over any finite field, but the perfect solution may not.
\end{Corollary}
\begin{IEEEproof}
The optimal solution exists as long as a solution exists, which is evident because a $K\times K$ identity matrix is a trivial solution over any finite field. We prove the second half via the example below, in which we show that uniform matroid $U^2_4$ is not binary representable, and thus the reception instance which has $\W=\I$ does not have a perfect binary solution according to Corollary 1.
\end{IEEEproof}

\begin{Example}
Uniform matroid $U^2_4$ has a ground set $E=\{\e_1$,$\e_2,\e_3,\e_4\}$ and rank 2. Its $\I$ is the collection of all two-element and one-element subsets of $E$. Representing it over $\F_2$ requires a $2\times 4$ binary matrix with 4 distinctive non-zero columns. Nevertheless, only 3 such columns are possible, as shown below. Hence, $U^2_4$ is not binary representable.
\begin{table}
\centering
\caption{$U_K^r$ is $q$-{representable} iff $K\leqslant K'$.}
\begin{tabular}{c|c|c}
$r$&$K'$&restriction on $q$\\\hline
1& no & no\\\hline
2&$q+1$& no\\\hline
\multirow{2}{*}{3}&$q+1$& $q$ odd\\\cline{2-3}
&$q+2$& $q$ even\\
\end{tabular}
\hspace{1em}
\begin{tabular}{c|c|c}
$r$&$K'$&restriction on $q$\\\hline
\multirow{2}{*}{4}&$5$& $q\leqslant 3$\\\cline{2-3}
&$q+1$&$q\geqslant 4$\\\hline
\multirow{2}{*}{5}&$6$&$q\geqslant 4$\\\cline{2-3}
&$q+1$&$q\geqslant 5$\\
\end{tabular}
\label{tab:uniform}
\end{table}
\begin{equation}\label{eq:u24}
\mC=
\begin{tabular}{c|cccc}
~&$\e_1$&$\e_2$&$\e_3$&$\e_4$\\\hline
$\c_1$& 0 & 1 & 1 & ?\\
$\c_2$& 1 & 0 & 1 & ?
\end{tabular}
\end{equation}

Hence, the reception instance that has $\W=\I$ does not have a perfect binary solution. Its optimal binary solution is a $3\times 4$ matrix and defines a matrix matroid with $\I\supset\W$.
\end{Example}

Results on the representability of uniform matroids are complete for $r\leqslant5$ and summarized in Table \ref{tab:uniform} \cite{Matroid_theory}. They serve as a useful reference for reception instances with $\wm\leqslant 5$ . However, there are no complete results for $r>5$.

A common approach to identify the representability of a general matroid is through \emph{forbidden minors} \cite{Matroid_theory}. A minor of a matroid is a smaller matroid obtained by element deletions and/or contractions on the original matroid. A matroid is $q$-representable iff it does not contain some forbidden minors. For example, $U^2_4$ is the only forbidden minor to identify binary representability. However, the complete lists of forbidden minors are only available for $q$ up to 4. Moreover, there are no polynomial-time algorithms to identify whether a matroid has a forbidden minor or not.

The above open problems in matroid theory hinder the identification of the perfect DLNC solutions. Furthermore, the collection of all Wants sets, $\W$, of a reception instance may not be sufficient to define a matroid when $\W$ is an incomplete collection of independent sets, i.e., when $\W\subset\I$.

Therefore, an algorithm which is able to generate a matrix matroid over any finite field based on any collection $\W$ of independent sets is highly valuable. We propose one such algorithm in the next section. It heuristically constructs a graphic matroid for any $\W$, and thus can directly complete the collection of independent sets. More importantly, this algorithm also directly produces DLNC solutions over any finite field because, according to Lemma \ref{lemma:graph_matrix}, the graphic matroid is representable over any finite field.

\section{Efficient Generation Algorithm}
In this section, we propose and then numerically evaluate our matrix matroid/DLNC solution generation algorithm.

Its core is to find a graphic matroid $M(\G)$ which has $\I\supseteq\W$ using $K$ iterations. Initially, the graph $\G$ has $U=2$ vertices: $\v_1$ and $\v_2$, and no edges. In the $k$-th iteration, we first list all the independent sets that contain $\p_k$ and subsets of $\{\p_1,\cdots,\p_{k-1}\}$ according to $\W$. We then list the forbidden locations of $\e_k$ which would otherwise introduce graphic circuits that violate the listed independent sets. Among the allowed locations, we choose the one with the smallest vertex indices\footnote{Defining a better criterion for such choice, as well as a better criterion for row removal in the matrix matroid, are our future research topics.}. If there is no allowed location, we add a new vertex, i.e., update $U\rightarrow U+1$, and allocate $\e_k$ to the location incident by $\v_1$ and $\v_U$, abbreviated as $\v_{(1,U)}$. Such allocations maximize the number of non-zero entries in the row representing $\v_1$ in the matrix matroid. Then by removing this row, the sparsity of the resulted DLNC solution is maximized, which could reduce decoding complexity. The complete algorithm is presented in Algorithm 1.

\begin{algorithm}[t]
\caption{~Graphic matroid based generation algorithm}\label{algorithm:graph}
\begin{algorithmic}[1]
\STATE \textbf{Inputs}: Wants sets $\W$, finite field size $q$.
\STATE \textbf{Initialization}: an empty index set $S$, a counter $k$, a graph $\G(V,E)$ with $U=2$ vertices and no edges.
\FOR {$k=1:K$}
\STATE According to $\W$, list all the independent sets between $\p_k$ and $\{\p_1\sim\p_{k-1}\}$. Denote their collection by $\I$ (\emph{Some sets in $\I$ might be subsets of others.});
\STATE For every independent set $I\in\mathcal I$, find the edge location of $\e_k$ that makes $I$ a circuit. Denote the collection of these forbidden edge locations by $F$;
\STATE Remove edge locations in $F$ from the collection of all the $U(U-1)/2$ possible edge locations. Denote the resulting collection by $F^*$;
\IF {$F^*$ is empty}
\STATE Add a new vertex to $\G$. Update $U\rightarrow U+1$. Allocate $\e_k$ to $\v_{(1,U)}$;
\ELSE
\STATE Allocate $\e_k$ to the location in $F^*$ with the smallest indices;
\ENDIF
\ENDFOR
\STATE Generate an all-zero $U\times K$ matrix $\mC$. For every column $\e_k$, let $c_{i,k}=1$ and $c_{j,k}=1$ if $\v_i$ and $\v_j$ are incident to $\e_k$ in the graph and $i<j$;
\STATE Remove the first row of $\mC$ and reduce $U$ by one. The resulting $\mC$ is the desired solution.
\end{algorithmic}
\end{algorithm}
\begin{Example}

Consider the SFM in \figref{fig:algorithm}(a). In the first iteration, $\e_1$ is allocated to location $\v_{(1,2)}$. In the second iteration, since $\p_2$ is independent to $\p_1$, the only edge location $\v_{(1,2)}$ is forbidden for $\e_2$. Thus we add a new vertex $\v_3$ and allocate $\e_2$ to $\v_{(1,3)}$. For the same reason, in the third iteration we add a new vertex $\v_4$ and allocate $\e_3$ to $\v_{(1,4)}$. In the fourth iteration, since $\p_4$ is independent to $\p_1$, $\p_2$, $\p_3$, and $\{\p_1,\p_3\}$ according to $W_{2,3,4}$, locations $\v_{(1,2)}$, $\v_{(1,3)}$, $\v_{(1,4)}$, $\v_{(2,4)}$ are forbidden, respectively. Then among the two allowed locations, we allocate $\e_4$ to the one with smaller indices, i.e., $\v_{(2,3)}$. In the fifth iteration, all the locations are forbidden and thus we add a new vertex $\v_5$ and allocate $\e_5$ to $\v_{(1,5)}$. The resulting graphic matroid is shown in \figref{fig:algorithm}(b) and its matrix matroid over $\F_q$ is shown in \figref{fig:algorithm}(c), which is a perfect solution because $U=\wm=4$.
\end{Example}

The algorithm finally outputs a $U\times K$ DLNC solution $\mC$. Comparing $U$ with $\wm$ results in four possible cases:

\noindent
\begin{itemize}
\item {Case-1}: $U=\wm$,  thus a perfect solution is found;
\end{itemize}
The remaining three cases take place when $U>\wm$:
\begin{itemize}
\item {Case-2}: The perfect solution exists, but the algorithm fails to find it;
\item {Case-3}: The perfect solution does not exist. In this case, if $U=\wm+1$ then $\mC$ is the optimal solution;
\item {Case-4}: $\mC$ is not optimal.
\end{itemize}
If the percentage that the algorithm hits {Case-1} overwhelms the others, and, among the rest, the percentage that $U=\wm+1$ dominates, we claim the algorithm is \emph{efficient}. The average of $U$ offered by an efficient algorithm is expected to be close to the average of $\wm$.

We verify our claim through extensive simulations. $K$ data packets are broadcast to $N$ receivers through wireless channels with i.i.d. packet erasure probability of $P_e$. The proposed algorithm is applied to the SFM after the systematic transmission phase. The resulting minimum number of coded transmission $U$ is compared with $\wm$, i.e., the lower bound.

For comparison, we also evaluate the minimum number of coded transmissions when RLNC over $\F_2$ and $\F_8$ is applied in the coded transmission phase, called systematic RLNC \cite{heide_systematic_RLNC}. The method is to keep producing random coefficient vectors until the rank of the $U\times K$ coding matrix reaches $\wm$.

The results under $K=15$, $N\in[5,40]$, and $P_e=0.2$ are presented in three figures. \figref{fig:average_U} compares the average of $U$ with the average of $\wm$. \figref{fig:perfect} and (c) display the percentages of the times that $U=\wm$ and $U\leqslant\wm+1$, respectively.
Our observations are:
\noindent
\begin{figure}
\centering
\subfigure[Average of $U$ compared with other schemes]{\includegraphics[width=0.8\linewidth]{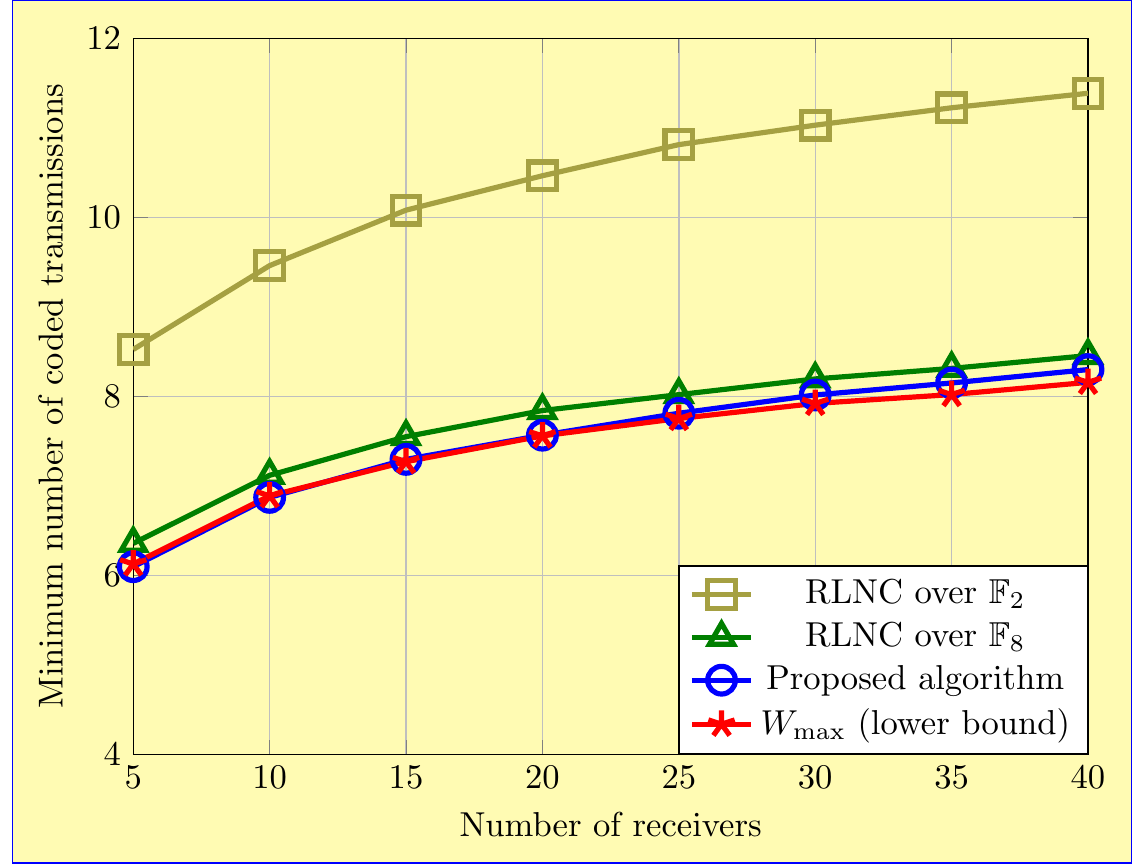}\label{fig:average_U}}
\subfigure[Percentage of $U=\wm$]{\includegraphics[width=0.45\linewidth]{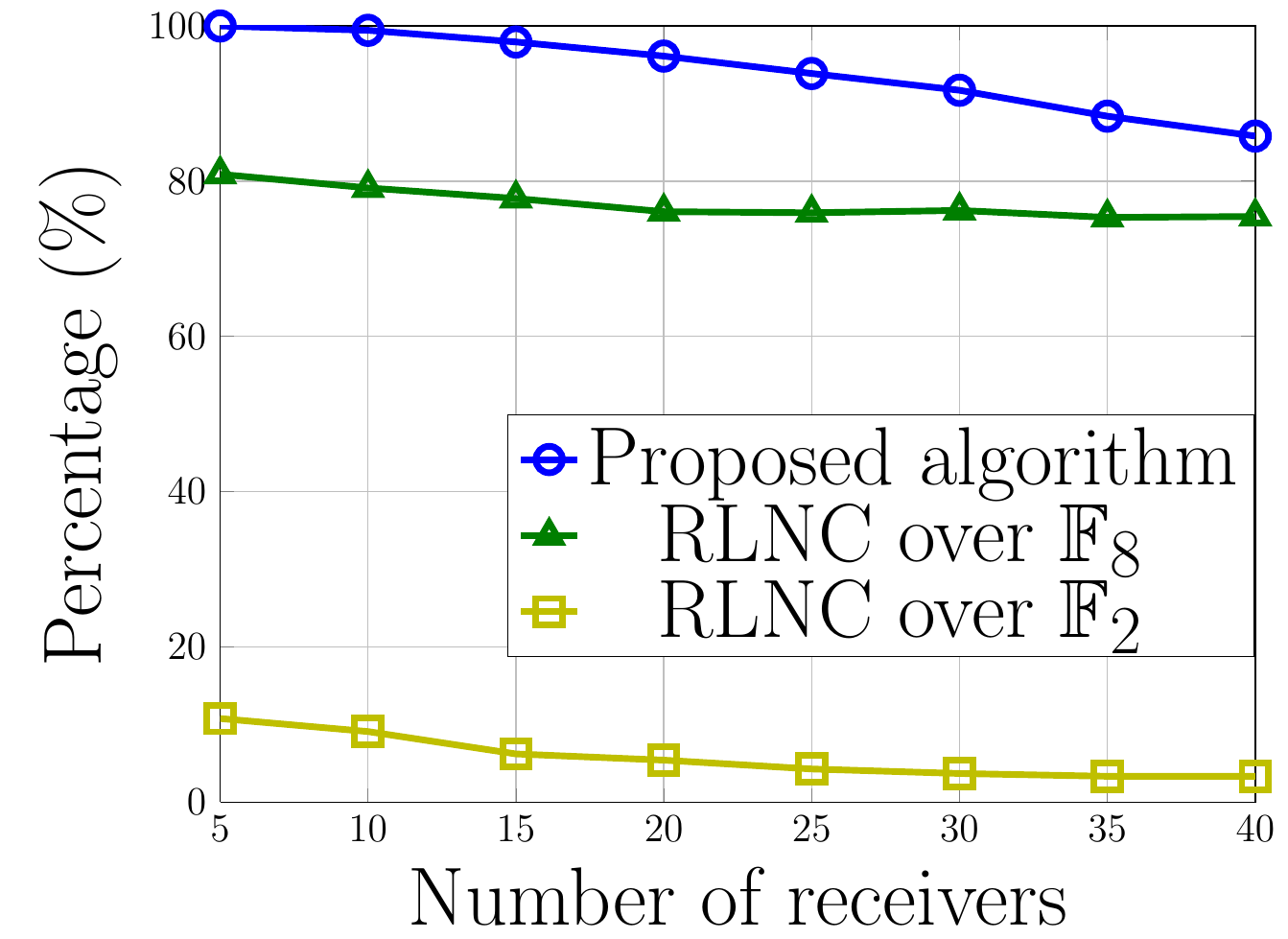}\label{fig:perfect}}
\subfigure[Percentage of $U\leqslant\wm+1$]{\includegraphics[width=0.47\linewidth]{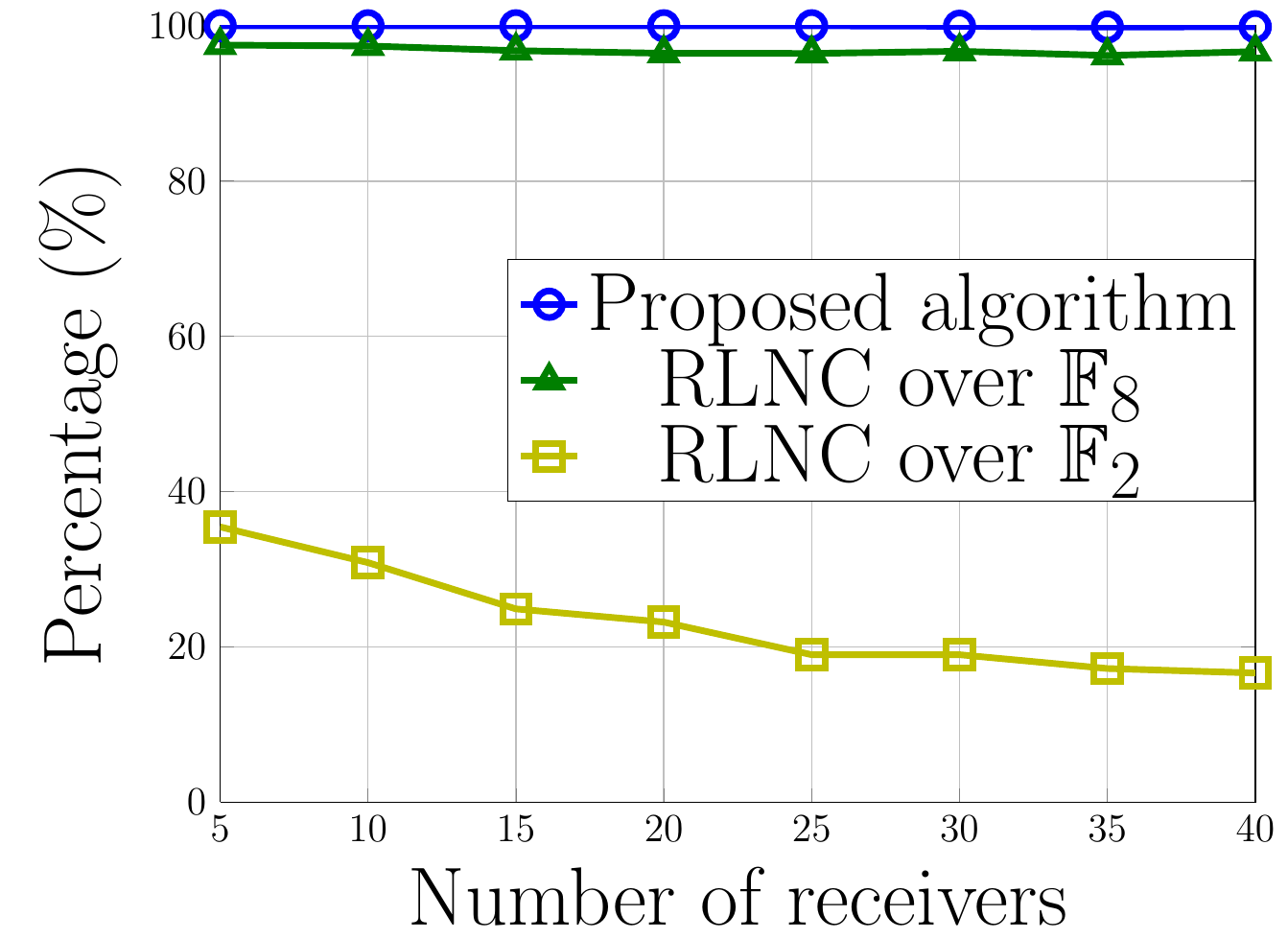}\label{fig:optimal}}
\caption{The performance of the proposed algorithm and RLNC with $K=15$ data packets and i.i.d. erasure channels with an erasure probability of $0.2$.}
\label{fig:performance}
\end{figure}
\begin{itemize}
\item The gap between the average $U$ of the proposed algorithm and the average $\wm$ is negligible for all the values of $N$. This is because it outputs perfect solutions for over $85\%$ of the time, as in \figref{fig:perfect}, and that it almost always provides $U\leqslant \wm+1$, as in \figref{fig:optimal}. Its performance only slightly degrades with increasing the number of receivers, possibly due to more complex $\W$.
\item The performance of the proposed algorithm is much better than its binary RLNC counterpart. The performance gap narrows down when the finite field increases to $\F_8$.
\end{itemize}
When the field size is sufficiently large, the performance of RLNC will coincide with the lower bound ($\wm$) and thus
 exceeds the proposed algorithm, but marginally, with the cost of heavy computational load.

\section{Conclusion}
In this paper, we employed matroid theory to model deterministic linear network coding (DLNC) in wireless packet broadcast scenario. Building upon on the equivalence between DLNC solution and matrix matroid, we studied the performance limits of DLNC in terms of the number of transmissions and its feasibility associated with the finite field size. Although a complete answer to its dual matroid representability problem is open, we still obtained some important results for DLNC, such as the sufficient and necessary condition for the existence of perfect DLNC solutions, and a graphic matroid based DLNC solution generation algorithm which works efficiently over any finite field.

This work also motivates many fundamental research opportunities, such as how to apply the rich results in matroid theory to further characterize DLNC, and how to design practical transmission schemes to broadcast the DLNC solutions.

\bibliographystyle{IEEEtran}
\bibliography{IEEEabrv,My_ref}

\end{document}